\documentclass[12pt]{article}
\usepackage{epsfig}
\begin{document}
\input{psfig}
\newcommand{\be}{\begin{equation}}
\newcommand{\ee}{\end{equation}}
\newcommand{\bq}{\begin{eqnarray}}
\newcommand{\eq}{\end{eqnarray}}


\thispagestyle{empty}
\setcounter{page}{1}


\vspace{0.88truein}
\centerline{\bf QUANTUM DYNAMICS OF COUPLED BOSONIC WELLS}
\vspace*{0.035truein}
\centerline{\bf WITHIN THE BOSE-HUBBARD PICTURE
}
\vspace*{0.37truein}
\centerline{\footnotesize Roberto Franzosi
,
Vittorio Penna
,}
\vspace*{0.015truein}
\centerline{\footnotesize\it
Dipartimento di Fisica \& Unit\`a INFM, Politecnico di
Torino} 
\baselineskip=10pt
\centerline{\footnotesize \it Corso Duca degli Abruzzi 24,
I-10129 Torino, Italy.} 
\vspace*{10pt}
\centerline{\normalsize and}
\vspace*{10pt}
\centerline{\footnotesize Riccardo Zecchina
}
\vspace*{0.015truein}
\centerline{\footnotesize\it Condensed Matter Section,
International Center of Theoretical Physics} 
\baselineskip=10pt
\centerline{\footnotesize \it
Strada Costiera 11, 34100 -Trieste, Italy} 
\vspace*{0.225truein}

\vspace*{0.21truein}
\abstract{
We relate the quantum dynamics of the Bose-Hubbard model (BHM)
to the semiclassical nonlinear equations that describe an array of
interacting Bose condensates by implementing a standard variational
procedure based on the coherent state method. We investigate the
dynamics of the two-site BHM from the purely quantum viewpoint by
recasting first the model within a spin picture and using then the
related dynamical algebra. The latter allows us to study thoroughly
the energy spectrum structure and to interpret quantally the
classical symmetries of the two-site dynamics. The energy
spectrum is also evaluated through various approximations
relying on the coherent state approach.}{}{}
\section{Introduction}
\vspace*{-0.5pt}
\noindent
The simplest way to describe an array of interacting bosonic wells is
represented by the second quantized Hamiltonian for charged bosons
\be
H_{_{BH}} =
\sum_{i} \left [ -\mu n_{i} +U n_{i} ( n_{i} -1) \right ] -
{{T}\over{2}} \sum_{\langle i,j\rangle} \left (a^{+}_{i} a_{j}+
a^{+}_{j} a_{i} \right ),
\label{BHM}
\ee
\noindent
where the operators $n_{i}\doteq a^+_{i} a_{i}$ count
the number of bosons at the lattice site $i$, and the destruction
(creation) operators $a_{i}$, ($a^+_{i}$) obey the canonical
commutation relations $[a_{i},a^+_{j}]=\delta_{ij}$.
The parameters $T$, $U>0$, and $\mu$ represent the hopping amplitude,
the strength of the Coulomb on-site repulsion, and the chemical
potential, respectively.

Hamiltonian (\ref{BHM}), known in the literature \cite{1} as the
Bose-Hubbard Model (BHM), was originally derived as the lattice
version of the second quantized $|\Psi|^4$--field theory used
to model superfluids adsorbed in porous media.
It describes in fact a gas of identical charges that hop through
a $D$ dimensional lattice, and exhibit a repulsive action
whenever two (or more) charges are situated at the same site.
The bosonic character of the lattice stands in the fact that, in
principle, an unlimited number of charges is allowed to occupy each
lattice sites. On this account the BHM is able to mimic appropriately
various features of systems such as granular superconductors,
short-length superconductors and Josephson junction arrays, hence
deserving a growing attention in the recent years (see references
in Ref. \cite{2}).
For the same reason the BHM seems to supply the appropriate
theoretic framework in which describing an array of interacting
Bose-Einstein condensates.

The recent years have been characterized by remarkable progresses
in constructing experimental devices able to confine some hundreds
of bosons so as to form real Bose-Einstein condensates (BEC)
\cite{3}. This has prompted as well the investigation of more
structured situations in which two condensates
(regarded as sites or wells filled by condensed bosons within
the BHM picture) can interact via tunneling processes~\cite{4}.
In particular, a quite rich scenery of dynamical behaviors
has been found to characterize the two-site case (named boson
Josephson junction) in Ref. \cite{5}, where two coupled
Gross-Pitaevskii equations for two condensate wavefunctions
are assumed as the minimal interaction scheme whereby
describing interwell processes. This represents however the most
general array of coupled BEC's so far realized even if BEC's with
complex architectures have been constructed~\cite{6}, and a
large amount of both experimental and theoretic work has been
devoted to construct arrays within the physics of uncondensed
ultracold atoms in optical lattices (see Refs. \cite{7},
\cite{8} and references therein).

Despite its almost purely theoretic value, it is interesting to
illustrate for conceptual reasons the link between the BHM and a
system of many bosonic wells regarded as interacting Bose condensates.
We reconstruct such a correspondence in Sec. 2 for the general case
and employ it in Sec. 3 to show how the two-site BHM
coincides, in a semiclassical picture, with a two-condensate system.  

In this paper we shall focus our attention on the quantum
dynamics of the two-site case. Classically, the problem is known
to be integrable since the dynamics of the complex site variables
$z_1$, $z_2$ required to account for the system state (these identify
with the wavefunctions of the two-condensate system) can be recast
in terms of two action-angle variables by exploiting the constant
of motion ${\cal N} \equiv |z_1|^2 +|z_2|^2$.
The reliability of the classical picture comes, of course, from the
fact that indeed the operators $n_{\ell}$ counts a large number of
particle and thus identifies with its semiclassical counterpart
$|z_{\ell}|^2$ via the correspondence scheme discussed in Sec. 2.

A smaller number of particles per site makes reasonable recovering
the quantum picture both because this is experimentally possible,
and because it is interesting to investigate thoroughly the
changes in the structure of the quantum states when going from
the pure quantum regime to situations with an intermediate,
more or less pronouced, semiclassical character.
Moreover, it is important to recall that the BHM exhibits a zero
temperature $\mu$-$T$ phase diagram which contains regions of finite
size (lobes) where the system loses its superfluid character becoming
an insulator~\cite{1,2}. In this respect we are interested in
making evident the cause of the presence of insulator regions in the
phase diagram in relation with the structure of the two-site model
eigenstates.

In Sec. 3 we review the semiclassical form of the two-site
dynamics and implement the standard requantization procedure
for comparing it to the spectrum provided by the exact quantum
treatment. In order to study the quantum dynamics of the two-site
BHM in Sec. 4 we represent the Hamiltonian within a spin algebra
realization based on the generators of su(2), in which the index
of the algebra representation is proportional to the total boson
number $N$.
This establishes the equivalence between the usual
description of the BHM Hilbert space through the bosonic Fock basis
and a simplified picture relying on the angular momentum eigenstates.
The latter, in turn, allows one to use the su(2) coherent states
(where the conservation of the total boson number has been
already incorporated) when performing the classical limit.

The spin picture also provides the basis for developing in Sec. 5
the single-boson picture of the system dynamics. This amounts to
reconstructing the exact dynamical algebra \cite{9} of the two-site
BHM. Explicitly, the model is reformulated within an enlarged algebra
--which identifies with su(N+1) in its fundamental representation--
where the two-site Hamiltonian takes the form of a many-boson
Hamiltonian on a (N+1)-site lattice whose population is formed
by a unique (effective) boson.
The algebraic formulation we introduce is viable to BHM's with more
than two site and provides a deeper insight of the dynamics complexity
from the group-theoretic standpoint. Moreover, it appears particularly
advantageous in studying the quantum counterpart of the
nonintegrable classic dynamics of the N-site BHM ($N>2$).
This problem will be considered elsewhere.  

At the operational level, such an approach entails a reformulation
of the quantum dynamics in terms of a (classical) Hamiltonian which
exhibits a quadratic dependence on the components of the su(2) picture
states. In particular, such an Hamiltonian is employed in Sec. 6 to
analyse various dynamical regimes and to recognize the distinctive
characters of both the superfluid and the insulator behavior.

%
\section{Arrays of condensates within the BHM picture}
\vspace*{-0.5pt}
\noindent
The link that relates the BHM to a system of interacting boson
condensates is easily established by implementing the method used
in Refs. \cite{2,10} for investigating the quantum dynamics of
many-particle lattice models.
To this end the one-dimensional (1D) Hamiltonian derived from
(\ref{BHM})
\be
H_c =\sum^M_{j=1} \left [ -\mu n_j +U n_j ( n_j -1) -
{{T}\over{2}} \left (a^{+}_j a_{j+1} + a^{+}_{j+1} a_j
\right ) \right ] \, ,
\label{HQ1}
\ee
suffices to generate a chain of $M$ interacting wells namely the
generalization of the case with two interacting condensates.

The method just mentioned combines the generalized
coherent state picture of quantum systems with the time-dependent
variational principle (TDVP) \cite{9}, and allows one to reformulate
the quantum dynamical problem relative to (\ref{HQ1})
by a semiclassical hamiltonian dynamics in terms of the expectation
values
\be
z_j := \langle Z| a_j |Z \rangle \; ,\;
z^*_j := \langle Z| a^+_j |Z \rangle
\label{EV}
\ee
and generated within the TDVP procedure by Hamiltonian
${\cal H}_c( Z, Z^*) $ $\doteq \langle Z | H_c | Z \rangle$.
The so-called trial macroscopic wave function $|Z \rangle$ here is
assumed to have the form $|Z \rangle = \prod_{j} |z_j \rangle$,
where $1 \le j\le M$ and $|z_j \rangle$ are the Glauber coherent
states \cite{9} solutions of the standard equation
$a_j |z_j \rangle = z_j \, | z_j\rangle$ for each $j$.
The dynamical equations for $z_{j}$ (those for $ z^*_j$ are
obviously obtained by complex conjugation)
\be
i \hbar {\dot z}_j = ( 2U |z_j|^2 - \mu) z_j - \frac{T}{2}
\sum_{i\in(J)} z_j
\label{EM1}
\ee
are derived~\cite{2} from the semiclassical Hamiltonian
\be
{\cal H}_c( Z, Z^*)
\equiv \sum_{j} \left [ (-\mu  + U |z_j|^2 ) |z_j|^2 -
{{T}\over{2}}
\left (z^*_j z_{j+1} + z^{*}_{j+1} z_j \right ) \right ] \, ,
\label{PIU}
\ee
through the standard formulas ${\dot z}_j =\{ z_j ,{\cal H}_c \}$
based on the canonical Poisson brackets $\{ z^*_j, z_{\ell} \}=
(i/\hbar)\delta_{j,\ell}$. It is easily checked as well that the
basic feature $[H_c, N] =0$ of the quantum system has a classical
counterpart given by $\{ {\cal H}_c , {\cal N} \} =0$, where
$ {\cal N} \equiv \sum_{\ell} |z_{\ell}|^2$.
The identification of $z_j$ with the wavefunction of the $j$-th
condensate establishes the link claimed above, showing
explicitly that the effective dynamics of BHM generates the classical
dynamics of coupled condensates.

\section{Semiclassical picture of the two-well system}

It is interesting to derive the two-well dynamics from the general
case of the $M$-site lattice described by Hamiltonian (\ref{PIU}).
Upon assuming $z_{2s} = z_2$ and $z_{2s+1} = z_1$ (suppose
that $M$ is even and the lattice is endowed with periodic boundary
condition), (\ref{PIU}) leads to the two-site Hamiltonian
\be
{\cal H} := {\cal H}_c/(M/2)
\equiv \left [ U( |z_1|^4 + |z_2|^4) -
\mu ( |z_1|^2 + |z_2|^2 )
- T \left (z_1^{*} z_2 + z_2^{*} z_1 \right )  \right ] \, ,
\label{TSE}
\ee
while (\ref{EM1}) reduce to the two equations 
\be
\cases{
&$ i \hbar {\dot z}_1 = ( 2U |z_1|^2 - \mu) z_1 - T z_2 $\cr
&${\-}$ \cr
&$ i \hbar {\dot z}_2 = ( 2U |z_2|^2 - \mu) z_2 - T z_1 $
\cr}
\; .
\label{EM3}
\ee
This amounts to considering a special solution characterized
by two independent complex variables with $|z_1|^2 +|z_2|^2$
representing the constant of motion. Remarkably, the same equations
are obtained independently when assuming ${\cal H}$ to be the
two-site Hamiltonian equipped with the Poisson brackets
$\{z_a, z_b^* \} = \delta_{ab}/i\hbar$, $a, b\, = 1, 2 $.
It is easily recognized that $q$-site models, where $q p= M$
for some $p\in {\bf N}$, can be achieved by following the same
procedure ($z_{\alpha +k q} \equiv \xi_{\alpha}$, $0 \le k \le p$,
$1 \le \alpha \le q$) and that their dynamics can always interpreted
as solutions of the $M$-site model exhibiting the discrete translation
symmetry $z_{j + kq} = z_j $ , $\forall j$, $0 \le k \le p$.
The physical interpretation of such a symmetry is given by the
Fourier modes description of the dynamical variables
${\sqrt M}\, b_q = \Sigma_j z_j {\rm exp}[i {\tilde q} j]$,
where ${\tilde q }:= 2\pi q/M$, $j, q \in [1, M]$, and
$M \delta (j-\ell)\equiv\Sigma_q {\rm exp}[i{\tilde q} (j-\ell)]$.
By inserting the condition $z_2= z_{2s}$ and $z_1 = z_{2s+1}$ one
finds $b_0 \equiv {\sqrt M}(z_1 + z_2)/2$, and
$b_{M/2} \equiv {\sqrt M}(z_1 - z_2)/2$ that imply
$$
z_1 = (b_{0} + b_{M/2})/ (2{\sqrt M}) \; ,\;
z_2 = (b_{0} - b_{M/2})/ (2{\sqrt M}) \; .
$$
Therefore the two-site model represents the simplest possible way
to construct an excited state with respect to the condensed
macroscopic state characterizing the zero temperature regime
where $z_j = b_0/{\sqrt M}$, $\forall j$.
Hamiltonian (\ref{TSE}) can be rewritten as
\be
{\cal H} \equiv \left [ \frac{U}{2}
{\cal N}^2 - \mu {\cal N} + \frac{U}{2} D^2
- T \sqrt{{\cal N}^2 - D^2} cos (2 \theta)  \right ] \, ,
\label{HS2}
\ee
where we have introduced the new canonical variables
${\cal N} = |z_1|^2 + |z_2|^2$, $D = |z_1|^2 - |z_2|^2$,
$\theta= (\phi_1 -\phi_2)/2$, and $\psi= (\phi_1 +\phi_2)/2$
implied by $z_j= |z_j|{\rm exp}(i\phi_j)$. The new
coordinates are equipped with the Poisson brackets
$$
\{\theta, {\cal N}  \} = 0 = \{ \psi, D \} \, , \,
\{\theta, D \} = -1/\hbar = \{ \psi, {\cal N} \} \, ,
$$
that provide the equation of motions
\be
{\dot D} =2T \sqrt{{\cal N}^2 -D^2} \, sin(2\theta)
\;  , \;
{\dot \theta} =-UD-\frac{TD}{\sqrt{{\cal N}^2 -D^2} }
\, cos(2\theta) \, ,
\label{EQ1}
\ee
the dot denotes the time derivative $d/d\tau$ with $\tau := t/\hbar$.  
Based on the complete integrability of the system, and proceeding
along the same lines of Ref. \cite{10}, one can reduce the above
two equations to a unique one
\be
{\dot D}^2 = -4 \left ( E -U{\cal N}^2/2 +
\mu {\cal N}  - U D^2/ 2 \right)^2 + 4T^2 ({\cal N}^2 -D^2) \, ,
\ee
where $E= {\cal H}$ is a given value of the energy, which
reproduces the result achieved in Ref. \cite{11}.
Going to its second order version
$$
{\ddot D}= \left ( a - b D^2 \right )D \, ,
$$
where $a =4U \epsilon - 4T^2$, $b=\,2U^2$,
and $\epsilon = E -U{\cal N}^2 /2 + \mu {\cal N}$ (reduced energy),
allows one to represent its solution in terms~\cite{12}
of the elliptic cosine $cn(x; k)$. This has the form
$D(t) = {\cal A} \, cn[\lambda (\tau - \tau_0); k]$,
in which ${\cal A}^2= (1+ a/\lambda^2)/(b \lambda^2)$,
$k= (1+ a/\lambda^2)/2 $, and $\lambda$ is a scale time that
can be fixed arbitrarily~\cite{12}.

In order to implement the requantization precedure a la
Brillouin-Einstein-Keller~\cite{13} the complete set of
the fixed point is requested. These are
\be
\theta = 0 \, , \, D=0 \, , \quad {\rm (minimum)}
\ee
\be
\theta = \pm \pi/2 \, , \,
D= \pm \sqrt{{\cal N}^2-(T/U)^2} \, , \quad {\rm (maxima)}
\label{MAXC}
\ee
\be
\theta = \pm \pi/2 \, , \, D=0 \, , \quad {\rm (saddle \, points\,
if \, T/U <{\cal N} )}
\ee
with (reduced) energy $\epsilon$
\be
\epsilon_m = -T{\cal N} \, , \quad
\epsilon_{_M} = \frac{U}{2} ( {\cal N}^2 + T^2/U^2) \, , \quad
\epsilon_S = +T{\cal N} \, ,
\label{ener}
\ee
respectively. The structure of the $D-\theta$ phase space 
depends on the parameter $\Gamma := T/{\cal N}U$.
For $\Gamma < 1/2$ the domains containing the maxima (consider,
for example, those related to $\theta = \pi/2$ represented in Fig. 1a)
are confined by a separatrix
\be
D_{M}^{(\nu)}(\theta)
= \nu {\cal N} \sqrt{ 1 - 4 \Gamma^2 cos^2(2\theta)}\, ,
\label{SEP}
\ee
with $\nu =+1$ ($\nu =-1$) for the upper (lower) maxima, that
corresponds to the energy $\epsilon_* = U {\cal N}^2/2 $. Such
disjoint domains are separated by a region
for whose orbits $\theta$ can increase indefinitely.
Another pair of separatrices confine the curves encircling the
minimum. These are obtained from
(\ref{HS2}) by setting the reduced energy $\epsilon$ equal to
$\epsilon_S$ and read
\be
D_{m}^{\pm}(\theta)= \pm \sqrt{2} {\cal N}
\left [ 1 - \Gamma \cos^2 (2 \theta)
\pm \cos (2 \theta)
\sqrt{ \Gamma^2 \cos^2 (2 \theta) + (1 -2\Gamma}) \right ]^{1/2}  .
\label{SEPA}
\ee
For $\Gamma = 1/2$ the separatrices of the two maxima intersect
in $\theta = \pi/2$, $D_{m}^{\pm}=0$ thus eliminating the possibility
of ballistic evolution of the phase. At the same time (see Fig. 1b)
they merge with the minimum separatrices (\ref{SEPA}).

For $\Gamma >1/2$ two-branch separatrix (\ref{SEP}) forms
a unique curve which is displaced vertically and join the upper
boundary of the phase space ($D_{M} = {\cal N}$) to the lower one
$(D_{M} =-{\cal N})$. The values in the interval
$1/2 < \Gamma <1$ correspond to situations in which
a new eight-shaped separatrix encircles the two maxima (see Fig. 2a)
and divides the curves of motion that locally sourround each maxima
from those that contain both.
Equation (\ref{SEPA}) still describes such a separatrix
centered in the saddle point ($D_{m}^{\pm}=0$, $\theta= \pi/2$),
but it is convenient to introduce the variable
$\phi :=\theta -\pi/2$ in view of the fact that
$(1 - 2 \Gamma)$ is now negative. In the limit
$\Gamma \to {(1/2)}^+$ the vertical separatrix and the
eight-shaped separatrix merge, while for $\Gamma \to 1^-$ the maxima
merge (see Fig. 2b) thus generating a configuration with a unique
maximum which characterizes $\Gamma \ge 1$.

An approximate evaluation of the energy level distribution
can be easily obtained for curves of motion close to
the extremal points just introduced. Upon expressing $D$
as a function of $\theta$
\be
D^{\pm}(\theta)= \nu {\cal N}
\frac{\sqrt{2}}{U} \left[ \frac{2\epsilon}{U {\cal N}^2}
- \Gamma^2 c^2(\theta)
\pm \Gamma c(\theta) \sqrt{
\Gamma^2 c^2(\theta) +1 -\frac{2\epsilon}{U {\cal N}^2}}
\right]^{\frac{1}{2}} ,
\label{CUR}
\ee
with $\nu =\pm 1$ and $c(\theta) = \cos (2 \theta)$, the
requantization procedure is enacted by setting
\be
\oint_\gamma \, D \, d \theta \equiv 2\pi (n+1/2 ) \, ,
\label{reqpro}
\ee
where $n\in {\bf N}$, $\gamma$ is some isoenergetic closed path,
and the constant $\hbar$ is missing since $D$ is a dimensionless
action variable. 

\begin{figure}[htbp]
{\centerline{\psfig{height=4.5cm,file=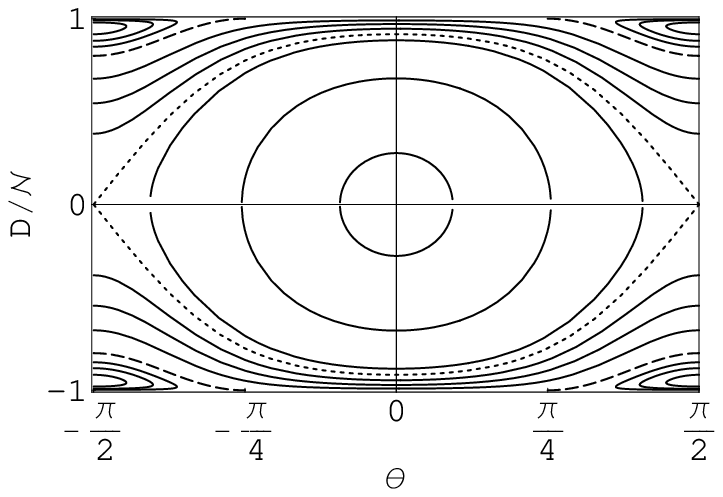}
\psfig{height=4.5cm,file=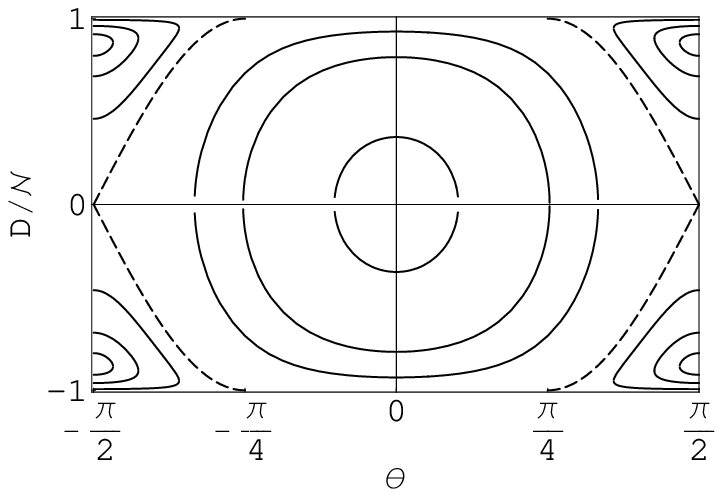}}}
\caption{
(a) left figure illustrates the minimum separatrix (dotted
lines) and the two separatrices (dashed lines) associated to each
maximum pair for $\Gamma < 1/2$; (b) right figure shows the merging
of separatrices for $\Gamma = 1/2$.}
\end{figure}

\begin{figure}[htbp]
{\centerline{\psfig{height=4.5cm,file=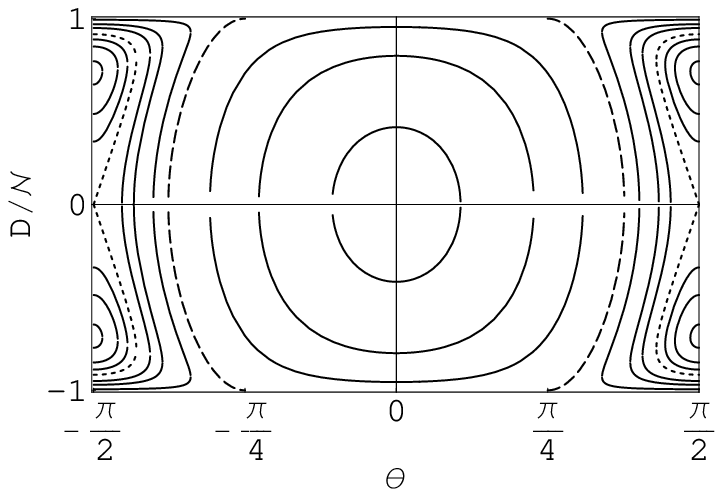}
\psfig{height=4.5cm,file=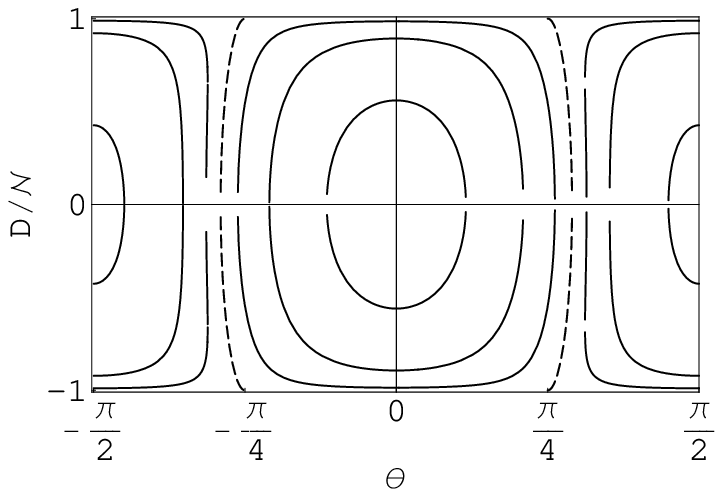}}}
\caption{(a) left figure illustrates the eight-shaped separatrix
encircling each maximum pair for $1/2 <\Gamma < 1$; (b) right figure
shows the phase space for $\Gamma > 1$. 
}
\end{figure}

In general, the exact calculation of the above
integral is quite difficult. Therefore it is convenient to reduce
$H$ to a diagonal quadratic form in proximity of both minima and
maxima. In the case of the minimum
$(D/{\cal N} \ll 1, \, \theta \ll 1)$ the resulting Hamiltonians reads
\be
{\cal H}_{m}  \simeq \frac{U{\cal N}^2}{2}  - {\cal N} (T + \mu)
+ \frac{U}{2}(1 +\Gamma ) D^2 + 2 T  {\cal N} \theta^2  \, ,
\label{HMIN}
\ee
while the maxima entail a quadratic Hamiltonian 
\be
{\cal H}_{M}  \simeq  U {\cal N}^2  -  \mu {\cal N} +
\frac{T^2}{2 U}
+ \frac{U}{2}[1 - \Gamma^{-2}]\, p^2 -2(T^2/U) \phi^2 \, ,
\label{HMAX}
\ee
with $p := [(D/{\cal N}) \pm (1-\Gamma^2)^{1/2}] \ll 1$,
$\phi \simeq \theta \pm \pi/2$.
Under such approximations the integral (\ref{reqpro}) can be easily
calculated and provides the approximate form of the spectra
\be
E_m(n) =  U {\cal N}^2 \left [ \frac{1}{2}
- \frac{\mu}{U{\cal N}} -\Gamma + 
\sqrt{ \Gamma (1 + \Gamma )} \, (2n+1) \right ]   
\label{EMIN}
\ee
and 
\be
E_M(n) = U {\cal N}^2 \left [1 - \frac{\mu}{U{\cal N}}
+ \frac{\Gamma^2}{2} - \sqrt{ 1 - \Gamma^2} \, (2n+1) \right ]  \, .
\label{EMAX}
\ee
for the minimun and the maxima, respectively.

The case when the Hamiltonian exhibits two maxima is quite
interesting in that the presence of pairs of
isoenergetic trajectories (spatially disjoint) suggests
the possibility for the system to undergo a tunneling effect
when considered quantum-mechanically.
Indeed this is the case for $ \epsilon_S < \epsilon < \epsilon_M$,
and $1/2< \Gamma < 1$ where isoenergetic trajectories appear
around the points
$$
P_R := (+{\cal N} \sqrt{1-\Gamma^2}, \pm \pi/2)
\, , \, \, P_{L} := (-{\cal N} \sqrt{1 -\Gamma^2}, \pm \pi/2) \, ,
$$
in the phase plane $(D, \theta)$. Due to the tunneling effect,
close to the value $\Gamma = 1/2$ the first order corrections
in $\hbar$ must be accounted when approximating the energy levels.

A simple way to evaluate such an effect is obtained via the following
standard scheme~\cite{14} showing how the tunneling probability
is nonvanishing when the barrier that separates the motions around
$P_R$ and $P_L$ is finite. In this case a new relevant contribution
(to the second order in the perturbative approximation) 
must be added to the energy levels (\ref{EMAX}).

Let ${H_R}$ and ${H_L}$ be the quadratic approximation of $H$
around $P_R$ and $P_L$, respectively, and
$| \Psi_R \rangle$ and $| \Psi_L \rangle$ the eigenstates
of ${H_R}$ and ${H_L}$ corresponding to the same eigenvalue
$E_M(n)$; the two eigenvalues:
\be
E_{\pm}(n) = E_M(n)  \, \mp \,
\hbar^2  \left [\Psi_R \, \frac{d \Psi_R}{dD} \right ]_C \, ,
\label{ELAND}
\ee
where $C$ is the crossing point $(D,\theta)=(0,\pi/2)$, correspond then
to the symmetric/antisymmetric states:
$| \Psi_{\pm} \rangle =(|\Psi_R \rangle \pm |\Psi_L \rangle)/ \sqrt{2}$.
At the second order we have
\be
\left [\Psi_R \, \frac{d \Psi_R}{dD} \right ]_C
= \frac{\omega}{2 \pi \hbar}
\exp \left [  -\frac{1}{\hbar} \int_{-a}^a d D \mid p(D) \mid
\right ] \, ,
\ee
where $\omega \equiv (U \sqrt{\Gamma^{-2} - 1})/ (2T) $,
$p(D) \equiv \left[2 \omega 
\left( \hbar n/2 - 2 \omega D^2/ U \right) \right ]^{1/2} $,
and the inversion points $D= \pm a$ of the motion along $D$  
have been issued from (\ref{CUR}) by setting $\theta= \pi/2$ in
$D^+(\theta)$.

\section{The spin picture of the quantum dimer}

The Hamiltonian of a lattice formed by two sites (dimer) reads
\bq
H = U(n^2_1 + n^2_2) - (\mu +U) N- T(a_1 a_2^+ + a_2a_1^+) \, ,
\label{HQ2}
\eq
where now $N \equiv n_1 + n_2$, whereas the factor $T$ of
the hopping term in place of $T/2$ is the only difference
distinguishing the open chain case from the case with the
periodic boundary conditions $a_{M+j} =a_j$ with $M=2$.
Upon introducing the two-boson realization
of the su(2) generators
\bq
J_1= \frac{1}{2} (a_1 a_2^+ + a_2a_1^+) \; ,\;
J_2= \frac{1}{2i} (a_1 a_2^+ - a_2a_1^+) \; ,\;
J_3= \frac{1}{2} (n_2 -  n_1)
\eq
which satisfy the commutators
$$
[J_1, J_2] = i J_3 \; , \; [J_2, J_3] = i J_1 \; , \;
[J_3, J_1] = i J_2 \; , \;
$$
and the Casimir equation
$C \doteq J_1^2 +J_2^2 +J_3^2 \equiv J_4 (J_4 +1)$
with $J_4= \frac{1}{2} (n_2 + n_1) $,  Hamiltonian (\ref{HQ2}) can
be rewritten in the nonlinear form
\bq
H = 2[ UJ_4^2 - (\mu +U) J_4 + UJ_3^2 - T J_1] \; .
\label{HJ2}
\eq
This fact has two consequences: First, the size of the Hilbert
space of the system, encoded in the eigenvalue $J$ of $J_4$, is
related explicitly to a macroscopic feature of the system since
$2 J_4$ coincides with the total particle number operator $N$.
The latter naturally plays the role of constant of motion within
our group-theoretic picture in that $J_4$ is a group central element
that is $[J_4, J_a] = 0$, $ a=1,2,3$.
Moreover, (\ref{HJ2}) shows how, due the nonlinear term $J_3^2$,
$H$ cannot be diagonalized via a simple unitary trasformation
of the group SU(2).
In view of the nonlinearity of (\ref{HJ2}) the problem of
diagonalizing $H$ for any value of $J$ can be faced either by
expressing $J_3^2$ (and $J_1$) linearly in terms of the generators
of a larger (dynamical) algebra, or by constructing approximation
procedures capable of retaining the relevant features of the spectrum.
The latter point of view is assumed in this Section.

Consider first the low part of the spectrum of $H$, the features of
which should be relevant to the zero temperature phase
diagram of the BH model. The spin formulation of the problem allows
one to implement an approximate approach based on the Casimir
operator. In fact, the observation that the classical energy minimum
is reached for $J_3= J_2 =0$, and $J_1 = +\sqrt C$, where
$C =J^2_1 +J^2_3 +J^2_2$ is a constant, leads to approximate $J_1$
as $J_1 \simeq {\sqrt C}[ 1 - (1/2C)(J^2_3 +J^2_2)]$ which implies
\be
H \simeq 2 \left \{ UJ_4^2 - (\mu +U) J_4 + UJ_3^2 -
T {\sqrt C}
\left [ 1 - \frac{1}{2 C} (J_3^2+ J^2_2) \right ] \right \} \; .
\label{HJA}
\ee
The variables $J_3 , \, J_2 $ (considered "small") account for small
displacements around the energy minimum.
At the quantum level, the inequalities
$\langle J_s \rangle \ll {\sqrt C}$ concerning
the expectation values of $J_s$, with $C= J(J+1)=N(N+2)/4 >> 1$,
and $s=2,3$ make natural to adopt the perturbative scheme where the
operator $J_2$ and $J_3$ can be treated as canonically conjugate
variables since
\be
 [J_2, J_3] \simeq\, + i \sqrt C \, ,
\label{COM}
\ee
whereas
\be
[J_3, J_1] = + i{\sqrt C} (J_2/ {\sqrt C}) \simeq 0 \; , \;
[J_2, J_1] = - i{\sqrt C} (J_3/ {\sqrt C}) \simeq 0 \, ,
\ee
are considered vanishing when compared with (\ref{COM}) since
$\langle J_s \rangle / \sqrt C << 1$. Then, after labelling by
the integer $p$ the eigenstates of the harmonic oscillator term
$[1 +T/(2U{\sqrt C})] [J_3^2+ W^2 J^2_2] $ contained in (\ref{HJA}),
where $W^2= [1+2U\sqrt C/T]^{-1}$, its eigenvalues are found to be
${\sqrt C} W (2p +1)$ so that the spectrum of $H$ reads
\be
E_p = \frac{U}{2} N^2 - (\mu +U)N
-2 T {\sqrt C} +
T \left [1+ \sqrt C \frac{2U}{T} \right ]^{\frac{1}{2}} \, (2p+1) \; .
\label{EIG1}
\ee
The approximation just implemented is of course exact for $UN =0$.
Hence its natural range of validity seems to be given by the condition
$T/NU \ge 1$, which can be interpreted semiclassically as the
requirement
that the expectation value of the hopping term $T|\langle J_1 \rangle|
\simeq TN/2$ is greater than the (maximum value of the) Coulomb
expectation value $U \langle J_3^2 \rangle \simeq UN^2/4$.
This appears to be consistent with the fact that the harmonic
oscillator form of the above spectrum is originated by the spectrum
of $J_1$ which is both linear and bounded from below.
When $T/NU <1$ the spin structure of the problem fully emerges since
$J_3^2$ introduce the influence of its quadratic degenerate spectrum. 
 
The importance of the condition $T/NU \ge 1$ is even more
evident when describing the spectrum close to the maximum
energy configuration. In this case the approximation
$J_1\simeq -{\sqrt C}[1-(1/2C)(J^2_3 +J^2_2)]$
makes appear in $H$ the harmonic oscillator term 
$-[T/(2U{\sqrt C}) -1] [J_3^2+ \Omega^2 J^2_2] $ 
where $\Omega^2= T/(T-2U\sqrt C)$ which involves a
spectrum of the form
\be
E_q = \frac{U}{2} N^2 - (\mu +U) N
+ 2T{\sqrt C} -
T \left [1-\sqrt C \frac{2U}{T} \right ]^{\frac{1}{2}} \, (2q+1) \; .
\label{EIG2}
\ee
For $T/NU < 1$ the system no loger exhibits a unique maximum
so that the oscillator approximation of the problem does not
work: the squared frequency $\Omega^2$ becomes imaginary. The
spin picture thus appears as the correct approach for understanding
quantum mechanically the degeneracy caused by the presence
of two (classical) symmetric maxima for $T/NU <1$.

\section{Two-well system dynamical algebra}

\subsection{Single-boson picture}
Considering the subalgebra su$_N$(2), in the spin representation
$J = N/2$ where $N$ is the total number of boson, within the algebra
$\cal A$ = su(M) allows us to recast the nonlinear Hamiltonian
$H$ in terms of a linear combination of generators
belonging to $\cal A$. The latter is therefore a {\it dynamical
algebra} for the system.
The simplest possible choice for $\cal A$ is provided by
the two-boson operator realization of su$_Q$(M) the generators of
which have the standard form
\be
E_{ij}:= b_i^+ b_j \, , \, (i \ne j) \, , \,
H_{ij}:= (b_i^+ b_i - b_j^+ b_j )/2
\label{GSU}
\ee
with $1 \le i, j \le M$, and satisfy the commutation relations
$$
[E_{ij}, H_{kl}] = \frac{1}{2} \left( \delta_{jk} E_{ik}\!
-\! \delta_{jl} E_{il}\! + \! \delta_{il} E_{lj}\!
-\! \delta_{ik} E_{kj}
\right) \; , \;
[E_{ij}, E_{kl}] = \delta_{jk} E_{il} \!
- \! \delta_{il} E_{kj} \, .
$$
The index Q eigenvalue of the invariant operator
$N_b =\sum_i b_i^+ b_i$ ($[N_b, E_{ij}]=0$) that selects the specific
representation of su(M) considered,
denotes the total number of particle associated with such a bosonic
realization of su(M).
In fact, the dimension of the Hilbert space basis ${\sl B}(M,Q)$ =
$\{ |n_1, ...,n_M \rangle, \sum^M_{i=1} n_i =Q \}$
is given by $dim\, {\sl B}(M,Q) = (Q +M-1)!/[(M-1)! Q!]$,
where the basis states are defined as
$|n_1, ...,n_M \rangle= \otimes^M_{i=1} |n_i \rangle $ and the
number operator states $|n_j \rangle$ fulfil the equations
$b_i|n_i \rangle = {\sqrt{n_i}} |n_i-1 \rangle$
and $b^+_i |n_i \rangle = {\sqrt{n_i +1}} |n_i+1 \rangle $.

Setting M=N+1 (recall that N is the boson number of the two-well
system) and Q=1 one selects the fundamental realization of
su(M), whose one-boson quantum states $|0, ...,1, 0... \rangle$
($dim\, B(N+1,1) \equiv N+1$) are in one-to-one
correspondence with the column vector standard basis whose matrix
form reads
$\{ |1 \rangle := (1, 0...)^T, |2 \rangle := (0, 1...)^T,...
|q \rangle := (0,..., 1, 0...)^T \} $.
The generators of the resulting matrix realization of
su$_1$(N+1) have the form
$$
||E_{ij}||_{qp} \equiv \delta_{qi} \delta_{jp} \,\, ,\,
||H_{ij}||_{qp} \equiv \frac{1}{2} (\delta_{qi} \delta_{ip} -
\delta_{qj} \delta_{jp}) \, ,
$$
where $1 \le q,p \le M=N+1 $. The states $|J; m \rangle$ constituting
the standard basis of su$_N$(2) can be similarly realized within the
column vector basis $\{|q \rangle,  1\le q \le N+1 \}$ via the
identification $|J; m \rangle \equiv |q \rangle$ with $m = J+ 1-q$.
The ensuing matrix version of the su$_N$(2) generators reads
$$
||J_+||_{q p} \equiv {\sqrt{(p-1)(2J+2-p)}} \delta_{q\; p\! -\!1} \; ,\;
||J_3||_{q p} \equiv (J-p+1) \delta_{qp}
$$
thus entailing that $J_+$ and $J_3$ can be rewritten within
su(M) (recall that $M \equiv N+1$) as
\be
J_+ = \sum^{N+1}_{q=1} {\sqrt{q (2J+1-q)}} E_{q\; q\!+\!1}
\, , \,
J_3 = \sum^{N+1}_{q =1}
\frac{J+1-q}{M} \left[ Q - 2\sum^{M}_{j =1}
H_{jq} \right] \, ,
\label{BOR}
\ee
respectively. Formally, this turns out to be equivalent to representing
our quantum description in the fully bosonic form, with
$$
J_+ = \sum^{N+1}_{q=1} {\sqrt{q (2J+1-q)}} b^+_q  b_{q+1}
\, , \,
J_3 = \sum^{N+1}_{q =1} (J+1-q) n_{q}  \, ,
$$
with the constraint $\sum^{N+1}_{q=1} b^+_q  b_{q} \equiv 1$.
Hamiltonian (\ref{HJ2}) then becomes
\be
H = C(J) +
2U \sum^{N+1}_{q =1} m^2(q)\, n_{q}
- T \sum^{N+1}_{q=1} R(q,J)\,(b^+_q b_{q+1} + b^+_{q+1} b_{q} ) \; ,
\label{HOB}
\ee
where $R(q,J) := [(J+1/2)^2-(m(q)-1/2)^2]^{1/2}$,
$m(q) := J+1-q$, and $C(J)=2 [UJ^2 -(\mu +U)J]$.
The related dynamics is now described by states
\be
|\Psi \rangle = \sum^{N+1}_{q =1} \xi_q  b^+_q |0,...,0 \rangle \, ,
\label{STA}
\ee
that,due to the normalization condition, turn out to define the
manifold $\sum^{N+1}_{q =1}\, |\xi_q|^2$ in the bosonic Hilbert space.
Such a manifold is an hypersphere in ${\bf C}^{N+1}$ and represents
the effective space in which the quantum dynamics of $H$ can be
reformulated in a classical form in such a way to mantain a complete
equivalence with the original quantum problem.
In fact the classical dynamics thus obtained actually concerns
the time evolution of the components of $|\Psi \rangle$ once
the initial condition  $|\Psi (0) \rangle$ at $t=0$ has been assigned.

Based on Eq. (\ref{STA}), the Schr\"odinger problem
$(i \partial_t - H) |\Psi \rangle =0$ can be
rewritten as a set of equations of motion
\be
i {\dot \Psi}_m = 2 U m^2 \, \Psi_m
-T \left [\rho_{_J} (m+1)\, \Psi_{m+1} + \rho_{_J} (m)\, \Psi_{m-1}
\right ] \, ,
\label{PSI}
\ee
where we have adopted the notation $\Psi_m := \xi_q$, and
$\rho_{_J} (m) := R(q,J)$ with $m := J+1-q$, in order to make the
present formalism closer to the spin picture introduced previously.
Equations (\ref{PSI}) can be derived in an independent way from the
effective Hamiltonian
\be
{\cal H}[\Psi] = C(J) + \sum^{N/2}_{m=-N/2}
\left [ 2U m^2\, |\Psi_m|^2
- T \rho_{_J} (m)
\left ( \Psi^*_m \Psi_{m-1} + \Psi^*_{m-1} \Psi_m \right )
\right ] \; ,
\label{HJB}
\ee
representing the energy expectation value
$\langle \Psi|H |\Psi \rangle \equiv {\cal H}[\Psi]$, provided the
Poisson structure $\{ \Psi_q,\Psi^*_p\}=\delta_{qp}/i \hbar$ is
assumed.
The eigenfrequencies associated to (\ref{HJB}) and the related
periodic orbits coincide with the eigenvalues and the eigenstates of
(\ref{PSI}), respectively. The latter are obtained from the
solution of the secular equation
\be
E X_m = 2 U m^2 \, X_m
-T \left [\rho_{_J} (m+1)\, X_{m+1} + \rho_{_J} (m)\, X_{m-1}
\right ] \, ,
\label{SEC}
\ee
that is easily derived from the equations of motion (\ref{PSI}).
The eigenstates components $X_m$ in Eqs. (\ref{SEC}) can be shown
to be real.

\subsection{Spectrum structure}

Two symmetries characterize the structure of the eigenstates. The
first is realized via the action of the unitary transformation
$U_1 := {\rm exp}[i \pi J_1]$ which takes $J_3$ into
$U_1 J_3 U^+_1 =-J_3$ so that $[H ,U_1 ]=0$. The action of $U_1$
on the standard basis is given by
$U_1 |m\rangle ={\rm exp}[i\pi J] |-m \rangle$.
This corresponds to implementing the component transformation
$X_m \to s X_{-m}$ with $s=\pm 1$ in Eqs. (\ref{SEC}) that remain
unchanged in that $\rho_{_J} (m+1) \equiv \rho_{_J} (-m)$ for each
$m$. Each eigenstate has therefore a definite symmetry character
which we make explicit by writing the eigenstates as
\be
|E\rangle_+ =\sum_m S_m (E) |m\rangle \, ,\, S_m := + S_{-m} \, , 
\label{ESF}
\ee
\be
|E\rangle_- =\sum_m A_m (E) |m\rangle \, ,\, A_m := - A_{-m} \, ,
\label{EAF}
\ee
in the symmetric and antisymmetric case, respectively.
The energy spectrum is not degenerate in that one has that
$E \ne E^{\prime}$ for any pair of states $|E \rangle_+$
and $|E^{\prime}\rangle_-$.

A further symmetry is obtained by combining the action
$U_3 J_1\, U^+_3=-J_1$ of $U_3 := {\rm exp}[i\pi J_3]$ on
$-TJ_1$ in $H$, with the change $T \to -T$ which restores the
initial form of $H$.
The corresponding change of the eigenstate components
$X_m \to {\rm exp}[i\pi m] X_m$ is easily derived from the fact
that the standard basis is acted on by $U_3$ according to
the formula $U_3 |m \rangle ={\rm exp}[i\pi m] |m \rangle$.
This does not make change Eqs. (\ref{SEC}) provided the sign of
$T$ is reversed as well.

An important feature of the eigenvalues structure is recognized
by comparing the secular equations for $|E \rangle_+$ and
$|E \rangle_-$. We first consider the case when $J$ is a half of
an odd integer. Then equations (\ref{SEC}) can be written in the
reduced form
\be
0= (2 U m^2-E) \, C_m
-T \left [\rho_{_J} (m+1)\, C_{m+1} + \rho_{_J} (m)\, C_{m-1}
\right ] \, ,
\label{SER}
\ee
with $C= A, S$, for $1/2 < m \le J$, while for $m =1/2$ one has 
\be
0= [2 U\,(1/2)^2 \, - E\, ] \, C_{1/2}
-T \left [\rho_{_J} (3/2)\, C_{3/2} +
\eta \, \rho_{_J} (1/2)\, C_{-1/2}
\right ] \, ,
\label{SPE}
\ee
where $\eta = -1\, (+1)$ must be taken in the antisymmetric
(symmetric) case. The remarkable consequence of the presence of
$\eta$ in (\ref{SPE}) is that one may distinguish the eigenvalues
associated with the symmetric states from those related to the
antisymmetric ones depending on the sign of $\eta$.
How this happens is clearly shown by the eigenvalues equation. The
latter is obtained iteratively via the recurrence equation
for the determinant of the associated secular equation
\be
{\sl d}[E, m]=(2Um^2- \, E)\, {\sl d}[E,\, m+1] -
T^2 \rho^2_{_J}(m)\, {\sl d}[E,\, m+2] \, ,
\label{REC}
\ee
which terminates with ${\sl d}[E,\, J] = 2U\,J^2- E$, starting from
\be
0 = (U/2- E+ \eta \, T \rho_{_J} (1/2))\,  {\sl d}[E,\,3/2] -
T^2 \rho^2_{_J}(1/2) \, {\sl d}[E,\, 5/2] \, .
\label{ESEC}
\ee
Developing (\ref{ESEC}) by means of (\ref{REC}) a $(J+1/2)$-th
degree equation for $E$ is obtained. For $\eta =0$ no difference
distinguishes the symmetric case from the antisymmetric case that
therefore exhibit pairs of eigenstates with the same eigenvalues.
As soon as $\eta$ is switched on each eigenvalues bifurcates
relatively to the two possibilities $\eta >0$ and $\eta <0$.
For small values of $T$ the leading order in (\ref{ESEC}) is
$T$, so that it reduces to
$$
0 =
(U/2- E+\, \eta T\,  \rho_{_J}(1/2))\,
\otimes^{J}_{m>1/2} (2Um^2-\, E)\, .
$$
There results that only the two lowest eigenvalues are distinguished
from those of the unperturbed case. Increasing $T$ allows one to
improve the resolution: for example, to the order $T^2$, the four
lowest eigenvalues are separated.

The same scheme holds when $J$ is integer ($m= 0, 1, 2...$) but
the hierarchy described above exhibits two special cases, for
$m= 0,1 \,$:
\be
0= -E \, C_0 - T \sigma \rho_{_J} (1)\, C_{1}  \, ,
\label{SERI}
\ee
\be
0= (2 U  \, -E) \, C_1
-T \left [\rho_{_J}(2)\, C_{2} + \sigma \, \rho_{_J}(0)\, C_{0}
\right ] \, ,
\label{SPEI}
\ee
with $C= A, S$, whereas Eqs. (\ref{SER}) account for $2 \le m \le J$.
The parameter $\sigma$ must be set equal to one in the symmetric case
($C= S$), while in the antisymmetric case ($C= A$) the expected
elimination of the component $A_0$ is realized by setting $\sigma =0$.
It follows that the Hilbert space dimension decreases while the secular
equation will have a degree diminished of one. Explicitly, one has
\be
0 = -E \, {\sl d}[E,\, 1] - \sigma^2 \rho_{_J}^2(1) T^2\,
{\sl d}[E,\, 2] \, ,
\label{ESEN}
\ee
which reduces to $0 = {\sl d}[E,\, 1]$ for $\sigma \to 0$. In the
symmetric case a $(J+1)$-th degree equation for $E$ issues from
(\ref{ESEN}) through the same formula (\ref{REC}).

For $T$ sufficiently small the energy levels appear to be structured
in doublets (see Fig. 3) that mimick the emergence of the energy
degeneracy actually reached only for $T=0$.
The eigenvalues of the doublet sector satisfy the inequality $E>TN$,
consistently with the role played classically by the saddle point
energy $\epsilon_{_S}$ of formula (\ref{ener}). Thus the doublet
number decreases when $T$ is increased (see Figs. 4, 5). 
This effects issues from the fact
that both $\eta$ and $\sigma$ are multiplied by $T$ in (\ref{ESEC})
and (\ref{ESEN}), respectively. The limit $T \to 0$ thus incorporates
the limits $\eta, \,\sigma \to 0$ whereby the degeneracy crops up
(each $E$ is associated to a pair $|E \rangle_{\pm} $).
The eigenstates generated (in the same limit)
$|E \rangle_{\pm} \equiv (|m \rangle \pm |-m\rangle)/ \sqrt{2}$
have the same self-energy $E(0) = C(J)+ 2Um^2$ of the noniteracting
model (competition among the $m$ modes switched off by $T=0$, that is,
no interwell dynamics).

The splitting of energy levels caused by $T \ne 0$ guarantees
the important feature that eigenstates are structured so as to have
components that are either symmetric or anti-symmetric with respect
to the inversion $m \to -m$.
A full degeneracy, in fact, should
allow for the occurrence of states inducing a strong localization
in proximity of those values of $\langle J_3 \rangle$ that classically
correspond to the energy peak positions.
In this case the eigenstates should exhibit a marked semiclassical
character that corresponds to states where either the components
with $m >0$ or those with $m< 0$ are strongly depressed in order
to permit the localization effect.

Despite the absence of states involving a stable localization
around one of the two energy peaks for energy values $E > T/U$
when $T/UN < 1$, the realization of localized states can be attained
by superpositions of the two states corresponding to the doublets.
Each such a pair of states is composed by a symmetric
eigenstate and an antisymmetric one with an energy gap $\Delta E$
which vanishes for $T \to 0$. The configurations where the system
is localized undergo a periodic revival (with a life time
proportional to $1/ \Delta E$) which ensues from the time
dependence of the eigenstate interference mechanism. 

\begin{figure}[htbp]
{\centerline{\psfig{height=5.1cm,width=10.0cm,file=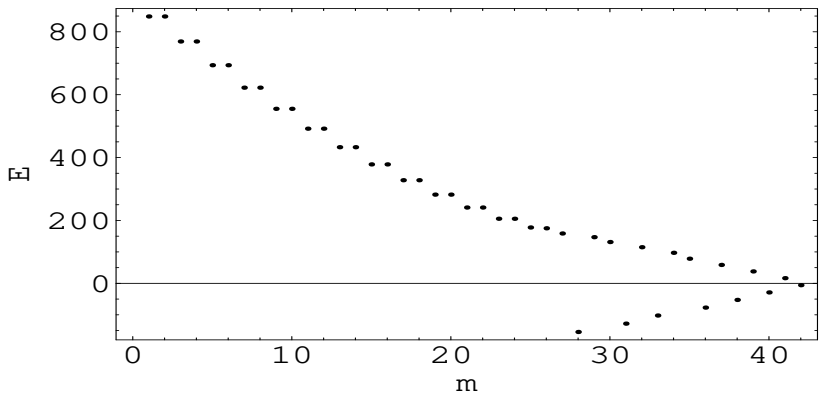}}}
\caption{Energy level distribution for $T=4$, $U = 1$, $N = 41$.
The index $m \in [1,42]$ labels the energy eigenvalues.}
\end{figure}
\begin{figure}[htbp]
{\centerline{\psfig{height=5.1cm,width=10.5cm,file=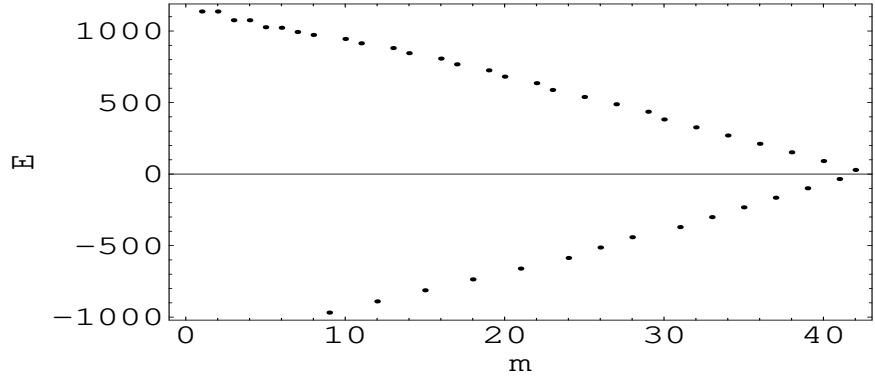}}}
\caption{Energy level distribution for $T= 24$, $U = 1$, $N = 41$.
The eigenvalues are labelled by the index $m \in [1,42]$.
Three doublets are still visible for $m <6$}
\end{figure}
\begin{figure}[htbp]
{\centerline{\psfig{height=5.1cm,width=10.5cm,file=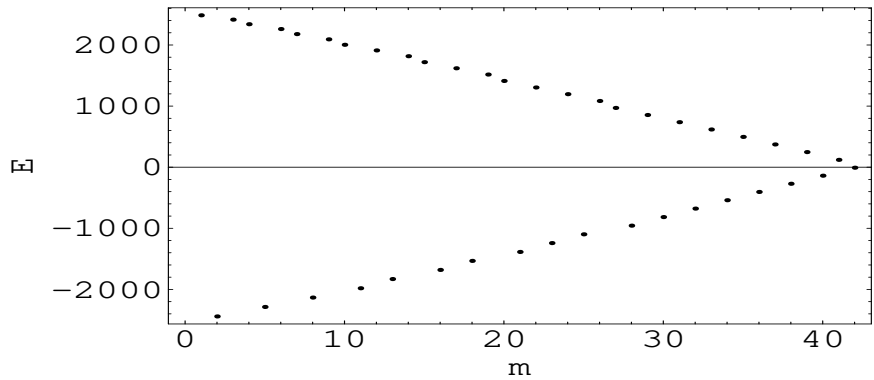}}}
\caption{$T=60$, $U = 1$, $N = 41$:
the influence of $U J^2_3$ in Hamiltonian (27) is almost negligible.}
\end{figure} 

\section{Discussion}

The structure of Hamiltonian (\ref{HJB}) allows us to interpret
the time evolution of a generic state $| \Psi \rangle$ as the
dynamics of an ensemble of linear oscillators with a space-dependent
coupling ($\rho_{_J} (m)$ depends on $m$) in the linear lattice of
$m$ modes. A certain amount of information can be extracted from
(\ref{HJB}) as to the form of the states $| \Psi \rangle$ close to
the extremal configurations.

Let us consider, first, the case $\Gamma \simeq 1$
(we identify the classical parameter $\cal N$ with the eigenvalue
$N =2J$ so that $\Gamma \equiv T/UN$).
The greatest coefficients of the hopping term
(${\rm max}[T \rho_{_J} (m)] \simeq U N^2/2$) are
comparable with the greatest coefficient of the Coulomb term
(${\rm max}[2U m^2] = U N^2/2$). On the other hand, such two classes
of terms involve different components that are labelled by
$m \simeq 0$ and $m \simeq \pm N/2$, respectively.
The minimum energy state is therefore characterized by components
$\Psi_m = R_m exp(i\alpha_m)$ such that $\alpha_m \equiv \alpha_{m+1}$
(the phase-locking involves the most negative the hopping term), whereas the modes
$|m| \ll N/2$ ($|m| \simeq N/2$) must be maximally occupied (strongly
depleted).
The dynamics of wealky excited states is expected to exhibit a similar
mode partecipation with the breaking of the (collective order of the)
phase-locking due to small phases' oscillations.
 
For opposite reasons the maximum energy state is characterized by an
antiferromagnetic type of phase order ($\alpha_{m+1} =\alpha_m +\pi$),
while both the largest modes ($|m| \simeq N/2$) and the smallest ones
($|m| \ll N/2$) contribute to maximize the energy. Such a phase order
enables us to recognize the largest components of the maximum energy
state starting from Hamiltonian (\ref{HJB}). Upon setting
$\alpha_m \simeq \pi +\alpha_{m-1}$ to maximize the energy, and
assuming that $||\Psi_m|-|\Psi_{m-1}|| \ll |\Psi_m|, |\Psi_{m-1}|$
which entails $\Psi^*_m \Psi_{m-1}+\Psi_m \Psi^*_{m-1} \simeq
-[|\Psi_m|^2+ |\Psi_{m-1}|^2]$, Hamiltonian (\ref{HJB})
reduces to the diagonal form
$$
{\cal H}[\Psi] \simeq C(J) + \sum^{N/2}_{m=-N/2}
\left [ 2U m^2\, + T \left ( \rho_{_J} (m)+\,\rho_{_J} (m+1)
\right ) \right ]|\Psi_m|^2  \, .
$$
For large values of $J=N/2$ the label $m$ can be regarded as a
continuous parameter so that the highest weight coefficient
$[ 2U m^2\, + T \left ( \rho_{_J} (m)+\,\rho_{_J} (m+1)
\right) ]$ is singled out by differentiation.
This provides $m\, [m \pm (N^2 -T^2/U^2 )^{1/2}/2 ] \simeq 0 $
where $m = \pm (N^2 -T^2/U^2 )^{1/2}/2 $ are the values for which
$|\Psi_m|$ must be maximized in order to reach
to the highest energy configuration. This result matches exactly
the classical representation of the maxima in the phase space
(see formula (\ref{MAXC})), except that quantum mechanically
both the maxima are involved in the dynamics.  
If $\Gamma \ge 1$, the solution $m= 0$ must be considered which
reflects in a consistent way the coalescence of the two (classical)
maxima in a unique one. 

The regime $\Gamma \ll 1$ is also intersting.
This case is quite sensitive to the parity of $N$ since the largest
Coulomb term has gained a factor N with respect to the largest
T-dependent term. If $N$ is even (i.e. $m$ assumes integer values)
the system ground-state must be constructed in such a way that
$|\Psi_0| \gg |\Psi_m|$ since $|\Psi_0|$ does not contribute to
the Coulomb term. We thus approximate $|\Psi \rangle$ by neglecting
any $\Psi_m$ such that $m \ne 0, \pm 1$. The resulting approximation
of ${\cal H}[\Psi]$ reads
$$
{\cal H}[\Psi] \simeq C(J) +4U |\Psi_1|^2 -
4T \rho_{_J}(0) \, |\Psi_1||\Psi_0|,
$$
where we have assumed $|\Psi_{+1}|\equiv |\Psi_{-1}|$,
and the constraint $2|\Psi_{+1}|^2+|\Psi_0|^2 =1$ holds.
The latter suggests to set $\sqrt{2} |\Psi_{+1}| = {\rm sin} \alpha$
and $|\Psi_{0}| = {\rm cos} \alpha$ in the above formula, which
in turn appears to be minimized by
${\rm tg} \alpha = \sqrt{2} T\rho_{_J}(0)/U \simeq TN/\sqrt{2} U$.
The ground-state energy thus reads
$$
E_{gs} = C(J) + \frac{U}{4}[1 - \sqrt{1+{\rm tg}^2 \alpha}] \, ,
$$
that reproduces the semiclassical value $E \simeq C(J) -TN$ for
$T/U \le 1/N$.
If $N$ is odd ($m = \pm 1/2, \pm 3/2, ... $) the Coulomb term
never vanishes, and the main contributions to ground-state come
from $\Psi_{-1/2}$ and $\Psi_{+1/2}$, so that the related energy
is approximated by
$$
{\cal H}[\Psi] \simeq C(J) +
\frac{U}{2} (|\Psi_{-1/2}|^2+ |\Psi_{+1/2}|^2)
-2T \rho_{_J}(-1/2) |\Psi_{-1/2}||\Psi_{+1/2}|\, .
$$
In view of the normalization $|\Psi_{-1/2}|^2+ |\Psi_{+1/2}|^2 = 1$,
it reduces to 
$$
E_{gs} \simeq C(J) + \frac{U}{2} -\frac{TN}{2} \sqrt{1+2/N} \, ,
$$
under the assumption $|\Psi_{-1/2}|=|\Psi_{+1/2}|$.

The characters distinguishing the even case and the odd case
suggest the interpretation of such situations in terms of
insulator and superfluid states, respectively. In the odd case
one has $m = \pm 1/2, \pm 3/2,...$ so that the ground-state is
the superposition of the states $|N/2, N/2+1\rangle$ and
$|N/2+1, N/2\rangle$ which involves a permanent one-boson tunneling
between the two wells (superfluid regime).
In the even case the ground-state essentially coincides with
$|J,0 \rangle \equiv |N/2, N/2 \rangle$ which entails a stationary
regime where the two bosonic wells are equally filled and the boson
tunnelling is not favoured (insulator regime). It is worth noting
that any contribution of the hopping term turns out to be definitely
smaller than the Coulomb terms namely
${\rm max}[T \rho_{_J}(m)] \simeq U/2 < {\rm min}[U m^2] \simeq U$
when one sets $\Gamma  \equiv 1/N^2$. The ensuing almost null
importance of the hopping term is precisely the signal of the
transition from the superfluid to the insulator regime which is
known (see \cite{1}, \cite{2}), for a given $N$, to take place
at $\Gamma \equiv 1/N^2$.

A simple evaluation of the components $X_m$ of the eigenstates
associated with the energy spectrum extremes (where relative sign
of $X_m$, $X_{m+1}$ is definite) is obtained by regarding $m$ as
a continuous parameter in Equation
(\ref{SEC}). In fact, upon setting
$X_m = Y_m/ \sqrt{(J +m)! (J -m)!}$
and replacing $Y_{m \pm 1}$ with $Y_{m} \pm dY_{m}/dm$
in (\ref{SEC}), one finds 
\be
0 = ( E- 2 U m^2 +2JT) \, Y_m -2T \sigma m \frac{dY_{m}}{dm}
\, ,
\label{SEX}
\ee
where $\sigma =+1 \, (-1)$ for the minimum (maximum) energy
state. The solution have the form $Y_m = C m^{\Lambda} exp(-Um^2/2T)$
with $\Lambda = E/(2T\sigma) -J$ and the consatant $C$ is fixed by
imposing the state normalization. The same treatment can be extended
to the
intermediate eigenstates but the sign oscillation of $X_m$'s, known
from numerical calculation of eigenstate, requires a more accurate
reformulation of Eqs. (\ref{SEC}) also involving a second derivative
term.

\section{Conclusions}
In this paper we have focused our attention on the quantum aspects
of the dynamics of a two-site Bose-Hubbard model. The latter mimicks
the interaction of two BEC's via tunneling effect. This has led us
to study thoroughly the nontrivial structure of the energy spectrum the
upper part of which consists of a series of doublets if $\Gamma <1 $,
and has suggested the quantal interpretation of the degeneracy
characterizing the classical configurations (a pair of symmetric
orbits is associated to the same value of the energy).

The general framework in which we relate the quantum dynamics
of BEC's to the classical formulation usually employed for
macroscopic condensates has been described in Sec. 2,
where the purely quantum model of many lattice sites (bosonic
wells coupled within the BHM picture) is reduced to a classical
lattice field theory by means of the TDVP procedure and the coherent
state representation of the macroscopic (quantum) state of the
lattice gas. Conversely, such a construction shows that a generic
array of coupled BEC is naturally reconducted to the standard
BHM and its generalization to nonperiodic lattice
structures \cite{15} when considered in a nonclassical regime
(small or mesoscopic number of bosons per condensate).

Sec. 3 has been devoted to illustrate the classical form of the
two-site model, its phase space, and the orbit structure therein.
After discussing the integrable character of the system the dynamics
of which is completely accounted by canonical variables $\theta$,
$D$ (see equations (\ref{EQ1})), the requantization
procedure {\it a la} BEK has been implemented.
This provides integral (\ref{reqpro}) whose approximate solutions
give the evaluation of energy levels (\ref{EMIN}) and (\ref{EMAX})
in proximity of the minimum and the maxima, respectively. The nice
feature that (\ref{EMAX}) holds for $\Gamma <1$ has allowed us to
incorporate the exchange effects via the procedure leading to
formula (\ref{ELAND}). Such a result anticipates the basic trait
of the spectrum represented by its doublet structure.

The reformulation of the two-site model in terms of a macroscopic
spin has been performed in Sec. 4. Such a picture offers the
possibility to approximate the energy spectrum in an independent
way for levels close to the minimum and maximum values. The
approximation well fits the exact energy values for $\Gamma >1$
and shows implicitly how the exhaustive investigation of the
spectrum unavoidably requires the spin picture. 
This paves the way for the construction of the dynamical algebra
which is the base for the exact diagonalization of the model.

The realization of $su_N(2)$ within the larger algebra
$su_Q(N+1)$ in terms of a single-boson representation (Q=1)
has been developed in Sec. 5 and supplies the spin operators
in the form (\ref{BOR}). It provides the algebraic framework
where the nonlinearity of $J_3^2$ is eliminated and the Hamiltonian
becomes a linear superposition (see (\ref{HOB})) of $su_1(N+1)$
generators. The procedure is easily generalized to many-site
lattice~\cite{16}.
The set-up developed leads to reformulate the two-site model
as a (N+1)-lattice model described by the quadratic Hamiltonian
(\ref{HOB}) and a population constituted by a single particle.
The spin nonlinear dynamics has been finally reconducted to a
classical problem with the quadratic Hamiltonian (\ref{HJB})
depending on the canonical variables $\Psi_j$ (these are the
components of the wavefunction solving the Schr\"odinger
equation) and $\Psi^*_j$. The proper modes of (\ref{HJB})
coincides with the eigenvalues $E$ of $H$ that are grouped
in two sectors $E >TN$ and $E<TN$ if $0< \Gamma <1$. The first
sector contains the doublets.

The basic result illustrated in Sec. 5 is that the levels
of each doublet are associated to a symmetric state and an
antisymmetric state exhibiting almost coincident secular equations.
Their difference, that turns out to be concentrated in one (or two)
of the equations forming the hierarchy of equations for the
eigenstate components (see Eqs. (\ref{SPE}) and
(\ref{SERI})-(\ref{SPEI})), has been represented via a parameter
which joins analytically the symmetric and antisymmetric eigenvalues.
This explains the origin of the doublet structure through a
bifurcation mechanism which represents the main result of the
present paper. In Sec. 6 the two site model ground-state has been
interpreted for $\Gamma \ll 1/2$ as a state with a superfluid
(insulator) behavior for an odd (even) total particle number $N$
thus showing how the germ of the lobe-like structure of the zero
temperature phase diagram is already present in a two-site lattice.
The method and the analysis developed in the present work can
be extended to lattices with a number of sites greater than two.
The bifurcation mechanism explained above is expected to provide
the base for studying the chaotic character of the trimer case.
Work on the three-site case is in progress along these lines~\cite{16}.
%

\vfill\eject
\noindent
{\bf Acknowledgements}
\vspace{0.3cm}

\noindent
The work of R. F. has been supported by an I.N.F.M. grant. V. P.
expresses his gratitude to the Condensed Matter Section of I.C.T.P.
for its hospitality as well as to the M.U.R.S.T. for financial support
within Project SINTESI.

\noindent

\end{document}